\documentclass[12pt]{article}
\usepackage{cite}
\usepackage{graphics}
\usepackage{epsfig}
\usepackage{graphicx}
\title{\LARGE\bf{Carbon films with a novel sp$^2$
network structure.\\}}

\author{\large\bf {I. Alexandrou$^{1,2}$, H.-J. Scheibe$^3$, C. J.
Kiely$^1$},\\ {\large\bf {A. J. Papworth$^{2,1}$, G. A. J.
Amaratunga$^{2,4}$ and B. Schultrich$^3$}}}

\begin{document}
\date{ }
\maketitle
\begin{center}
$^1$ Department of Materials Science and Engineering, University of
Liverpool,\\
 Liverpool L69 3BX, UK.\\
$^2$ Department of Electrical Engineering and Electronics, University of
Liverpool,\\
 Liverpool L69 3BX, UK.\\
$^3$ Fraunhofer Institute of Material and Beam Technology, D - 01277
Dresden, Germany.\\
$^4$ Department of Engineering, Cambridge University, Cambridge CB2 1PZ, UK.
\vspace{0.5cm}
\end{center}

\renewcommand{\baselinestretch}{1.5}
\small\normalsize

\begin{abstract}
Laser-Arc evaporation of a
graphite target has been used to deposit carbon films that exhibit high
hardness (45 GPa) and elastic recovery (85$\%$).
High Resolution Electron Microscopy (HREM) and Electron
Energy Loss Spectroscopy (EELS) were subsequently used to study the
microstructure
and bonding of the resultant layers. The structure of the films from
HREM is seen to consist of a dense array of
parallel curved graphene sheet segments packed in various orientations.
EELS reveals
that the films are comprised of mainly sp$^2$ bonded carbon.
The results suggest that a new form of carbon
thin film with fullerene-like structure can be realised.
In order to explain how a predominantly
sp$^2$ bonded material can exhibit such a high hardness,
a simple model is proposed to correlate the
excellent mechanical properties with the observed structure.

\end{abstract}
\section{Introduction}
~
The production of hard carbon based materials using cathodic
arc~\cite{330}
and laser ablation~\cite{307} techniques has been reported previously.
High film hardness has been attributed to the presence of a high
percentage of sp$^3$ (diamond~-~like) bonds,
whereas a high concentration of sp$^2$ (graphitic) bonds is
regarded as leading to the formation of soft films.
However, the discovery of the C$_{60}$ fullerene
molecule~\cite{203,198} and carbon nanotubes~\cite{327} which are sp$^2$
bonded, opens up the possibility of obtaining 3-D structures which
exploit the extremely strong~\cite{224} in-planar bonds of graphite
(stronger than diamond) in a new class of hard thin film materials.
The reports of the formation of hard sp$^2$ bonded materials by
anisotropically pressing C$_{60}$~\cite{196,197} and by embedding
distorted
fullerene~-~like nanoparticles in an amorphous carbon
matrix~\cite{nature} were early indications that such a carbon material
could be synthesised.
However, the structure of such high fraction sp$^2$ bonded materials
with superior mechanical properties is not known in detail.
In this paper we report a novel microstructure of hard and elastic carbon
films~\cite{emag97} prepared via a laser initiated pulsed cathodic arc
method (Laser-Arc).
Both High Resolution Electron Microscopy (HREM) and
Electron Energy Loss Spectroscopy (EELS) show evidence for
a new form of carbon thin film material which consists of sets
of curved graphene sheets mixed with amorphous carbon in which sp$^2$
carbon bonding dominates. Although a curved graphene sheet structure has
previously
been proposed for CN$_x$ materials~\cite{174,bernard}, this is the
first time that HREM and EELS have been used together in order to get a
more
complete view of both structure and bonding in a pure carbon material.

\section{Mechanical Properties}
~
The hard and elastic films were prepared using a carbon plasma produced
by the Laser-Arc method without any gas ambient~\cite{304}. Films
deposited by this method on Si substrates typically have Youngs
modulus values of $E = $ 400~-~700~GPa as measured using laser induced
acoustic waves~\cite{scheibe}. The value depends on the specific
deposition conditions (e.g. substrate temperature and arc current). For
the sample under investigation a Youngs modulus of E = 480 GPa was
determined. A nominal hardness of H = 45 GPa and an elastic recovery of
85$\%$ has been measured from indentation experiments. A typical
microindentation curve from such a film is shown in
Figure~\ref{lsrarc-hard}. The plastic
hardness quoted was calculated on the basis of the well known Olivier
and Pharr method~\cite{342,343} from
the indentation curves. There is some doubt regarding the absolute
accuracy of the hardness obtained by this method. Nevertheless, the
hardness is consistent with the rule of thumb H $\approx$ E / 10, which
is well proven for amorphous carbon films. The very
distinct characteristic of these films is the very small overall
indentation depth and the very high elastic recovery after being
subjected to a maximum load of the magnitude shown in
Fig.~\ref{lsrarc-hard}. This is a clear indication of a minimal plastic
deformation regime. Most other thin films (other than diamond) will show
strong evidence of plastic deformation at such load
levels~\cite{bull}. On  the other hand, a soft and elastic film, for
example rubber, will show a very high degree of elastic recovery, but
the maximum indentation depth would be an order of magnitude larger.
Therefore, indentation characteristics such as those measured from
Fig.~\ref{lsrarc-hard} are a clear signature of thin films with
superior mechanical properties.

\section{Microstructural characterisation}
~
HREM studies were carried out using a JEOL 2000EX microscope having
a 0.21~nm point to point resolution, whereas a VG
HB601 STEM with a 1~nm probe diameter was employed to collect the EELS
spectra.
The HREM and EELS analyses were performed on samples
prepared by cleaving the specimen perpendicular to the interface.
No preparation treatments involving ion beams or chemical etching were
used. This significantly reduced the possibility of introducing
microstructural artifacts.

Figures~\ref{lsrarc-tem} (a) and (b), show cross-sectional HREM
micrographs taken under differing defocus conditions from the same
thin area of the specimen. They reveal a structure which is very
different from that usually seen in amorphous carbon films.
Parallel curved fringes grouped together in sets are mixed forming
patterns of swirls and concentric rings, of typically 4-6~nm in
outer diameter. Selected area diffraction (SAD) showed two broad
diffuse rings at around 0.113~nm and 0.210~nm, which arise from
the amorphous carbon. In addition, a much sharper ring was seen
centred at 0.363~nm but with an actual spread between 0.358 and
0.368~nm. These values were determined using the Si substrate
material as calibration for the SADP. The 0.363~nm spacing is
consistent with the calculated spacing between the carbon layers
in bucky onion structures~\cite{388}. The shape of the fringes
suggests that they represent curved graphene sheets similar to
those found in carbon nanotubes and bucky
onions~\cite{nature,emag97}. These images seem to suggest that a
substantial part of the film is made up from fragments of such
nanoparticles mixed in a random way forming a co-continuous matrix
with the amorphous carbon component. As the specimen thickness
increases beyond 25~nm, it becomes progressively more difficult to
see the individual graphene sheets and the film takes on a classic
``amorphous" appearance. The characteristic pattern formed by the
graphene sheets is seen much more clearly in
Fig.~\ref{lsrarc-tem}~(b) which is taken close to Gaussian focus,
as compared to Fig.~\ref{lsrarc-tem}~(a) which was taken close to
Scherzer defocus. The details of the image contrast in these HREM
images are controlled by the contrast transfer function of the
microscope. The frequencies of interest are 2.75~nm$^{-1}$ which
corresponds to the graphene plane spacing (0.363~nm) and the
region around 4.76~nm$^{-1}$ which correspond to the range of
atomic spacings about 0.21~nm which are the periodicities
associated with the diffuse diffraction ring often observed in
amorphous carbon (a-C). This latter 0.21~nm spacing is close to
the (100) graphite and $\{111\}$ diamond spacings and can be
associated with either type of short range order in a-C.
Calculations of the contrast transfer function of our microscope
have been performed for Scherzer and Gaussian focus conditions. In
both situations the 0.363~nm (graphene) periodicity is faithfully
transferred. However, in the latter case the amorphous
periodicities around 0.21~nm are heavily attenuated relative to
that for Scherzer defocus. Consequently, for the image at Gaussian
focus, Fig.~\ref{lsrarc-tem}~(b), the graphene-like features are
accentuated at the expense of the amorphous component.

It is also possible to accentuate the graphene periodicities in the
Scherzer defocus micrograph using image processing techniques. For
example, the Fast Fourier Transform (FFT)
of the area indicated in Fig.~\ref{lsrarc-tem}~(a) has been filtered to
remove low periodicities and
subsequently back transformed to produce the clearer image shown in
Fig.~\ref{lsrarc-tem}~(c). Once again the curved graphene plane features
become strongly visible at the expense of the amorphous background.

Electron Energy Loss Spectroscopy was chosen as the most appropriate
tool for studying the bonding configuration in the
carbon network. Since this technique has been used extensively
for carbon films in the past~\cite{328}, correlation between our film
and other hard carbon materials reported to-date are possible.
Figure~\ref{lsrarc-eels} shows
EELS spectra acquired from different parts of the film.
The $1s-\sigma^*$ peak at 292~eV is the signature
of $\sigma$-bonding while the $1s-\pi^*$ peak at 285~eV indicates $\pi$
bonding in our material. The ratio of the integrals under the two peaks
is frequently used to estimate
the amount of $\pi$ bonding present~\cite{328}. Effects due to
background, plural scattering and zero loss energy width have been
deconvoluted from all these spectra. The result of these effects needs
to be
carefully considered to obtain an accurate measure of the $1s-\pi^*$
energy loss. We used a C$_{60}$ fullerite crystal as the
calibration material since it has a known 1:3 ratio of $\pi$ to $\sigma$
bonds and is free from the orientational effects which can arise in
graphite~\cite{yuan-emag}. Equating the heights of the $1s-\sigma^*$
peaks for all spectra, we can gain qualitative information about the
fraction of $\pi$-bonding in
our material by directly comparing the intensity of the $1s-\pi^*$
peaks.

Spectrum \ref{lsrarc-eels}(a) shows the C K-edge of a pure C$_{60}$
fullerite crystal which is
taken as the representation of a purely sp$^2$ material with randomised
bond orientation. Spectrum \ref{lsrarc-eels}(b) shows the C K-edge from
a highly
oriented graphite sample with the basal planes parallel to the
incident beam direction
(i.e. the $p$ orbitals are perpendicular to the optical axis). This
spectrum is used as a reference for the maximum orientational
enhancement of
the $1s-\pi^*$ peak that could occur for the convergence and collection
angles employed in this study~\cite{yuan-emag}.
Comparing spectra \ref{lsrarc-eels}(a) and \ref{lsrarc-eels}(b) we
see that although both materials studied consist of sp$^2$ bonded
carbon, the orientation of the basal planes in the latter case strongly
enhances
the $1s-\pi^*$ peak. Spectrum \ref{lsrarc-eels}(c) shows the C~K-edge
obtained from a
thin specimen area of our hard and elastic carbon film. Notably, the
intensity of the $1s-\pi^*$ peak lies in-between
the intensities of the reference materials, revealing the presence of at
least some orientational effects. The structure of our films shown in
Figs.~\ref{lsrarc-tem}(a) and (b) suggests that the 1~nm EELS probe is,
over the range of
a few curved graphene sheets, most likely aligned parallel to the
optical axis. This gives rise to orientational effects similar to those
seen in spectrum \ref{lsrarc-eels}(b). However, our material
is slightly different from the graphite sample described above because
within the cone defined by the convergent electron probe there are also
$\pi$ bonds with
various orientations due to the sheet curvature and the presence of
residual
amorphous carbon. Therefore, the $1s-\pi^*$ peak is not quite as high as
for the pure graphite sample. Following this line of argument, it would
also be expected that as film thickness increases and more
fullerene~-~like
patches overlap, an eventual averaging of $\pi$ bond
orientation should occur. Indeed, the intensity of the $1s-\pi^*$ peak
does decrease when we obtain the EELS spectrum from a thicker area of the
specimen, as shown in Fig.~\ref{lsrarc-eels}(d). The fact that
the $1s-\pi^*$ peak intensity actually decreases to a level slightly
below that for the C$_{60}$
fullerite is attributed to the presence of amorphous carbon with some
residual sp$^3$ bonding, the relative influence of which increases in
thick areas.

\section{Discussion}
\begin{figure}[tbp]
\begin{center}
\rotatebox{270}{\resizebox{!}{12.5cm}{\includegraphics{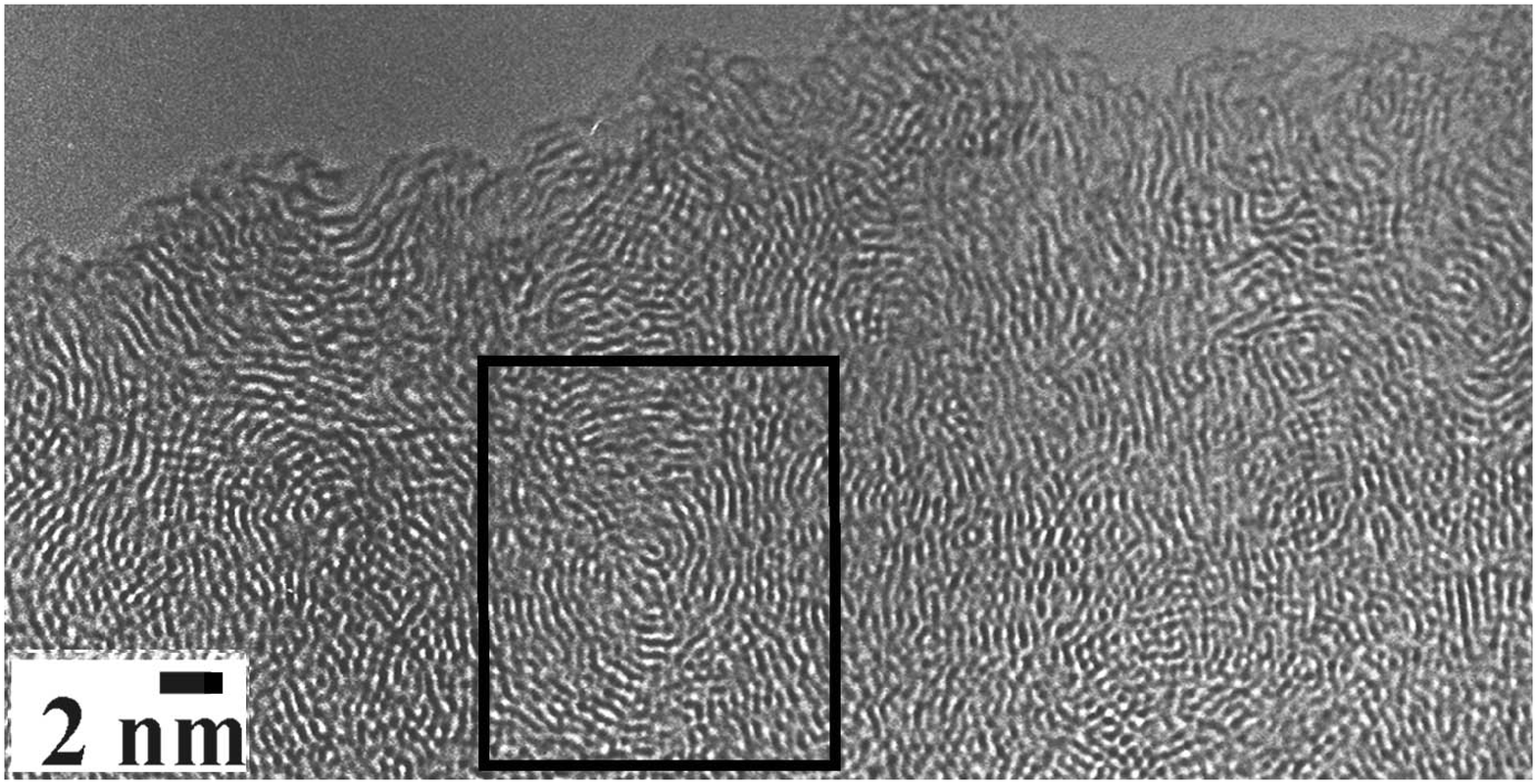}}}
\end{center}
\begin{center}
{\bf Figure 2(a)}
\end{center}
\begin{center}
\rotatebox{270}{\resizebox{!}{12.5cm}{\includegraphics{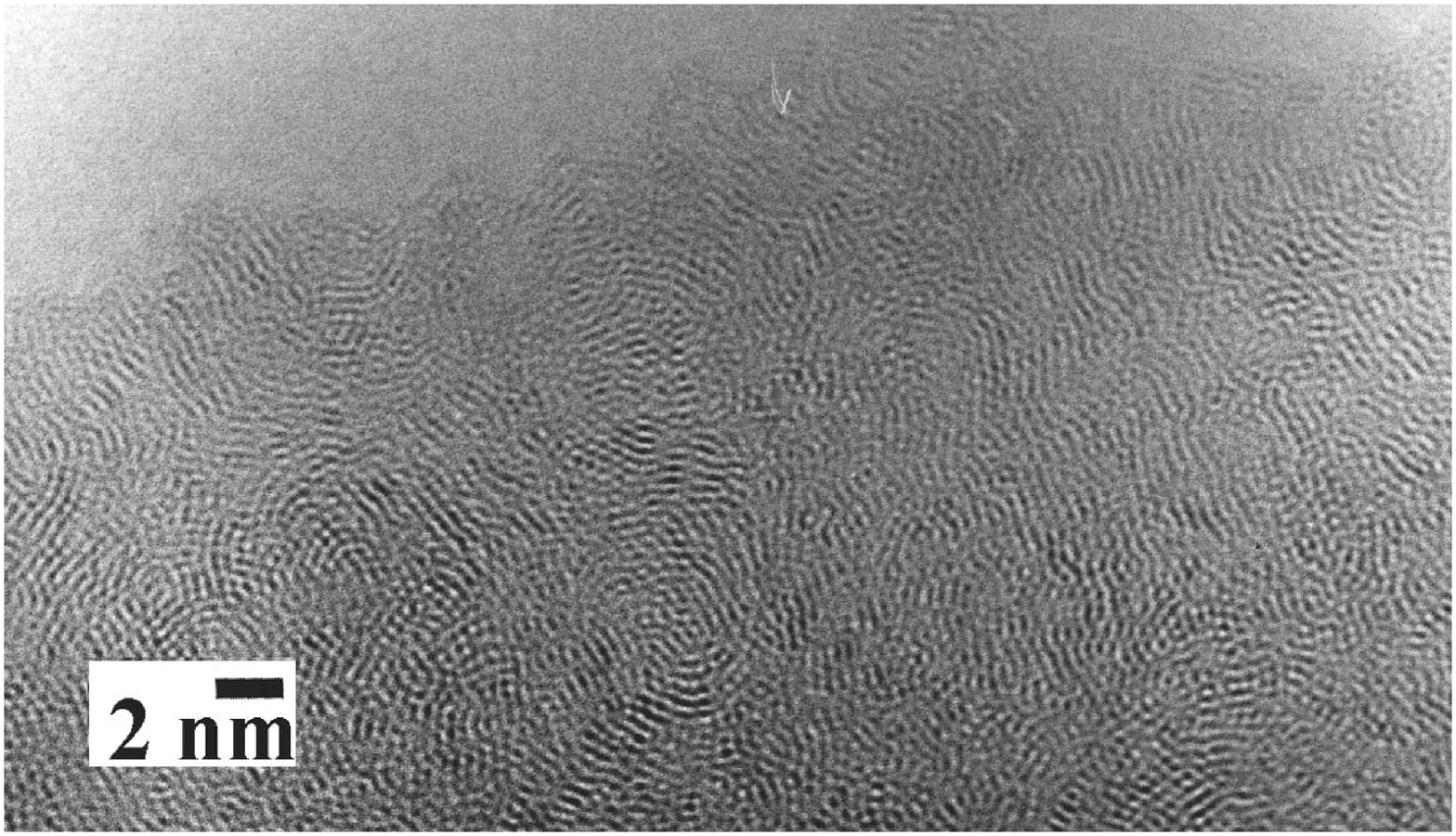}}}
\end{center}
\begin{center}
{\bf Figure 2(b)}
\end{center}
\label{lsrarc-tem}
\end{figure}

~ Since our material consists mainly of sp$^2$ bonded carbon, the
challenge is to relate the excellent mechanical properties with
the observed film microstructure. A ``squeezed chicken wire''
model is proposed to describe the observed network of graphene
sheets shown in Fig.~\ref{lsrarc-tem}. If pieces of flat chicken
wire are squeezed together so that they deform, they also become
linked because single wires coming off their edges are entangled
with the rest of the material. As a result it is very difficult to
separate them afterwards and a robust but yet flexible structure
can be formed. In analogy to pieces of chicken wire, a substantial
fraction of our material consists of curved graphene sheets. These
fullerene-like regions are brought into close proximity during
deposition, as is evident from the high density with which they
appear in HREM images. At their edges there are unsatisfied bonds
through which they may bond either to the amorphous material or to
other sheets. There have also been previous studies and
models~\cite{396,397} which have proposed that cross-linking of
graphitic sp$^2$ plane regions with occasional sp$^3$ ``defects"
can explain the optical and electronic properties of a-C films.
Here we have experimentally realised a hard graphitic material.
However, our structure is distinct from the previous models in
that it is pentagonal and heptagonal defects in the graphitic
plane which gives rise to curvature, and hence a pseudo-3D
graphitic structure by interlinking of these curved segments. This
is close to the structure envisaged by Townsend {\it et
al}~\cite{329} who on the basis of atomic modeling, suggested that
interlinking in purely sp$^2$ bonded structure can take place
through randomly oriented pentagonal, hexagonal and heptagonal
rings. Additionally, the carbon material reported here has a more
periodic structure than in a purely amorphous material, and the
curved graphene planes (fullerene-like structure) which lead to
the interlinking are clearly seen in the HREM image of
Fig.~\ref{lsrarc-tem}. Our material is best described as a
nanostructured carbon with a fullerene-like structure.

Unlike graphite which has strong in-plane covalent sp$^2$ bonds
and weak interplanar Van~-~der Waals bonds, the structure in our
hard and elastic films creates a 3-dimensional sp$^2$ bonded
carbon network. In a previous study~\cite{nature} of hard and
elastic carbon films resulting from fragmentation of carbon
nanoparticles, the current authors inferred that sp$^3$
diamond-like bonding was dominant in regions where nanoparticle
fragments interlinked. This deduction was based on the observation
of a reduced 1s-$\pi^*$ peak intensity in the EELS K-edge spectrum
obtained from an interlinked region, compared to that from
adjacent single nanoparticle regions. In light of the more
detailed EELS study carried out here, and taking into
consideration the orientational effects on the relative magnitude
of the 1s-$\pi^*$ peak, those earlier results may also be
interpreted according to the ``squeezed chicken wire" model. In
the earlier case, the orientational effects present in individual
nanoparticle fragments may have become randomised when they were
``squeezed" together in the interlinking region.

The
hardness of our films (45~GPa) is an order of magnitude higher than that
for graphite-like films ($\approx$~5~GPa). In a very simplified model,
our material may be considered as a strongly connected
arrangement of nanometer stacks of graphene lamellae with different
orientations. Depending on their orientation, the elastic modulus of a
single stack is given by the modulus in direction $C_{11}$ = 1060~GPa and
in the $c$ direction, $C_{33}$ = 36.5~GPa, respectively~\cite{extra3}.
Considering that
these differently orientated stacks usually represent parts of the same
(often closed) shell package the effective modulus may be approximated
by their parallel arrangement, i. e. as the arithmetic average. For an
isotropic film structure this gives $E_{eff} \approx (2 C_{11} + C_{33})/3
\approx 720$~GPa,
whereas for a textured structure $E_{eff} \approx (C_{11} + C_{33})/2
\approx 550$~GPa may be
expected. If the additional reduction by the surrounding amorphous
matrix is taken into account the latter value is consistent with the
experimental value E = 480~GPa. Hence, this crude estimation shows that
very high stiffness may be realised by suitably arranged graphene sheets
and it supports the impression of a certain degree of texturing.
Furthermore, when sets of graphene
planes are compressed their built-in curvature causes them to recover a
shape close to
their initial state after deformation. This we propose is the origin of
the apparently high elastic recovery in this material.

A structure that consists of a dense array of curved graphene sheets,
like the one seen in our films, is very close to the realisation of a
continuous fullerene-like carbon material.
Curved graphene sheets are well known for turbostratic structures. The
decisive difference in this case is the special arrangement of the
graphene lamellae representing closed or highly curved shells with
nanometer curvatures. In this way the relative gliding of the sheets is
prevented and extreme elastic recovery is possible by local buckling of
the sheets.
We propose that carbon films which exhibit this type of structure
constitute a new class of carbonaceous material.\\

{\bf {Acknowledgements :}} We would like to thank M. Johansson and L.
Hultman from Link$\ddot{o}$ping University for their expert guidance in
TEM sample preparation and A. Burrows for help with image processing. We
would also like to thank H. Ziegele for assistance with film deposition
and D. Schneider for the measurements of the Youngs moduli of the films.
Finally, we are indebted to Multi-Arc UK for funding this research
programme.

\newpage
\bibliographystyle{prsty2}
\bibliography{bib}

\begin{thebibliography}{10}

\bibitem{330}
D. R.McKenzie, D. Muller, and B.~A. Pailthorpe, Physical Review Letters {\bf
  67},  773   (1991).

\bibitem{307}
D.~H. Lowndes, D.~B. Geohegan, A.~A. Puretzky, D.~P. Norton, and C.~M. Rouleau,
  Science {\bf 273},  898  (1996).

\bibitem{203}
H.~W. Kroto, J.~R. Heath, S.~C. O'Brien, R.~F. Curl, and R.~E. Smalley, Nature
  {\bf 318},  162  (1985).

\bibitem{198}
W. Kr$\ddot{a}$tschmer, L. Lamp, K. Fostiropoulos, and D. Hoffmann, Nature {\bf
  347},  354  (1990).

\bibitem{327}
S. Iijima, Nature {\bf 354},  56  (1991).

\bibitem{224}
M.~M.~J. Treacy, T.~W. Ebbesen, and J.~M. Gibson, Nature {\bf 381},  678
  (1996).

\bibitem{196}
M.~E. Kozlov, M. Hirabayashi, K. Nozaki, M. Tokumoto, and H. Ihara, Applied
  Physics Letters {\bf 66},  1199  (1995).

\bibitem{197}
M.~N. Regueiro, P. Monceau, and J.-L. Hodeau, Nature {\bf 355},  237  (1992).

\bibitem{nature}
G.~A.~J. Amaratunga, M. Chhowalla, C.~J. Kiely, I. Alexandrou, R. Aharonov, and
  R.~W. Devenish, Nature {\bf 383},  321  (1996).

\bibitem{emag97}
I. Alexandrou, C.~J. Kiely, I. Zergioti, M. Chhowalla, H.-J. Scheibe, and
  G.~A.~J. Amaratunga, Institute of Physics Conference Series {\bf 153},  581
  (1997).

\bibitem{174}
H. Sj$\ddot{o}$strom, S. Stafstr$\ddot{o}$m, M. Boman, and J.-E. Sundgren,
  Physical Review Letters {\bf 75},  1336  (1995).

\bibitem{bernard}
B.~F. Coll, J.~E. Jaskie, J.~L. Markham, E.~P. Menu, and A.~A. Talin,  in {\em
  Amorphous Carbon : State Of The Art}, edited by S.~R.~P. Silva, J. Robertson,
  W.~I. Milne, and G.~A.~J. Amaratunga (Word Scientific, London, 1998), pp.\
  91--116.

\bibitem{304}
H.-J. Scheibe, B. Schultrich, H. Ziegele, and P. Siemroth, IEEE Transactions on
  Plasma Science {\bf 25},  685  (1997).

\bibitem{scheibe}
H.-J. Scheibe, D. Schneider, B. Schultrich, C.-F. Meyer, and H. Ziegele,  in
  {\em Amorphous Carbon : State Of The Art}, edited by S.~R.~P. Silva, J.
  Robertson, W.~I. Milne, and G.~A.~J. Amaratunga (Word Scientific, London,
  1998), pp.\ 252--261.

\bibitem{342}
W.~C. Oliver and G.~M. Pharr, Journal of Materials Research {\bf 7},  1564
  (1992).

\bibitem{343}
M.~F. Doerner and W.~D. Nix, Journal of Materials Research {\bf 1},  601
  (1986).

\bibitem{bull}
S.~J. Bull and S.~V. Hainsworth,  in {\em Amorphous Carbon : State Of The Art},
  edited by S.~R.~P. Silva, J. Robertson, W.~I. Milne, and G.~A.~J. Amaratunga
  (Word Scientific, London, 1998), pp.\ 272--278.

\bibitem{388}
Y. Yosida, Fullerene Science $\&$ Technology {\bf 1},  55  (1993).

\bibitem{328}
S.~D. Berger, D.~R. McKenzie, and P.~J.~S. Martin, Philosophical Magazine
  Letters {\bf 57},  285  (1988).

\bibitem{yuan-emag}
J. Yuan, N. Menon, G.~A.~J. Amaratunga, M. Chhowalla, and C.~J. Kiely,
  Institute of Physics Conference Series {\bf 153},  159  (1997).

\bibitem{396}
J.~C. Angus and F. Jansen, Journal of Vacuum Science and Technology A {\bf 6},
  1778  (1988).

\bibitem{397}
M.~A. Tamor and C.~H. Wu, Journal of Applied Physics {\bf 67},  1007  (1990).

\bibitem{329}
S.~J. Townsend, T.~J. Lenosky, D.~A. Muller, C.~S. Nichols, and V. Elser,
  Physical Review Letters {\bf 69},  921  (1992).

\bibitem{extra3}
E. Fitzer, Carbon {\bf 27},  621  (1989).

\end{thebibliography}
\newpage
\begin{center}
{\LARGE {\bf {Figure Captions}}} \vspace{1cm}
\end{center}

\begin{figure} [tbp]
\rotatebox{270}{\resizebox{!}{14.0cm}{\includegraphics{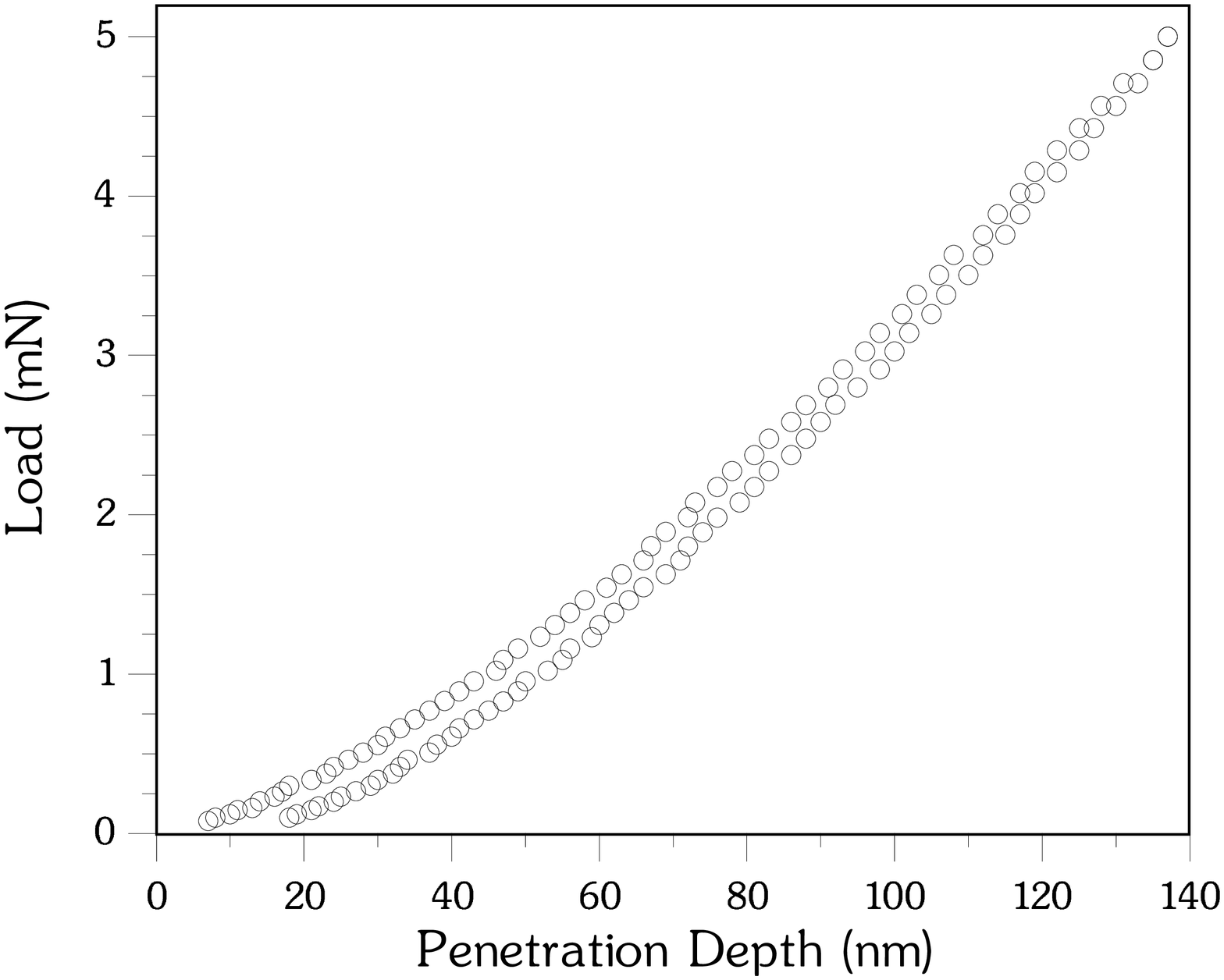}}}
 \label{lsrarc-hard}
 \vspace{2.0cm}
 \begin{center}
 {\bf Figure 1}
 \end{center}
\end{figure}

\begin {center}
{\bf {Figure~\ref{lsrarc-hard}}}
\end {center}

Typical load-displacement curves obtained during microindentation
testing. The elastic recovery is calculated using the formula
$\frac{d_{max}-d{min}} {d_{max}}$, where d$_{max}$ and d$_{min}$
are the maximum and minimum displacements during unloading,
respectively.

\begin {center}
{\bf {Figure~\ref{lsrarc-tem}}}
\end {center}

HREM micrographs : images $(a)$ and $(b)$ were obtained from the
same region of the film taken near Scherzer defocus ($\Delta f
\approx 52$~nm) and Gaussian focus ($\Delta f \approx 0$)
respectively. Both images show sets of parallel graphene sheets
forming swirls and concentric rings; image $(c)$-the area
indicated in image $(a)$ has been image processed to accentuate
the curved fullerene structure.

\begin{figure}[tbp]
\begin{center}
\rotatebox{270}{\resizebox{!}{12.5cm}{\includegraphics{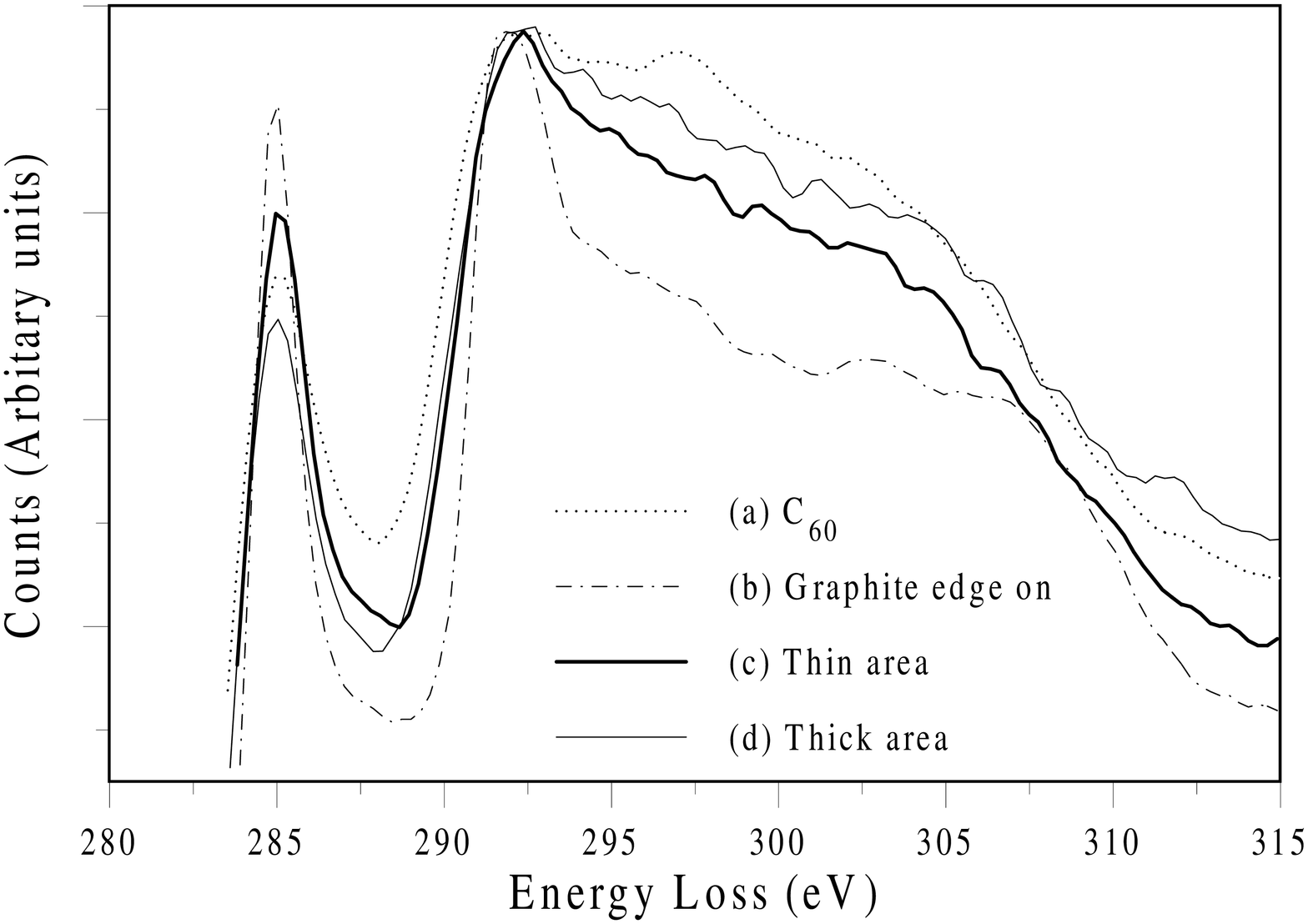}}}
\vspace{1.0cm}
\begin{center}
{\bf Figure 3}
\end{center}
 \label{lsrarc-eels}
\end{center}
\end{figure}

\begin {center}
{\bf {Figure~\ref{lsrarc-eels} :}}
\end {center}

EELS spectra obtained from :$(a)$ a pure C$_{60}$ fullerite
crystal which is regarded as the best description of a pure sp$^2$
material with an averaged $\pi$ bond orientation; $(b)$ a highly
oriented graphite sample with the basal planes parallel to the
optical axis; $(c)$ our hard and elastic carbon film where the
specimen has a thickness of 0.22$\times \lambda$; $(d)$ our hard
and elastic carbon film where the specimen thickness is
1.52$\times \lambda$. For all EELS spectra the energy resolution
was 0.3~eV/channel and the convergence and collection angles were
21.3 and 3.4~mrad, respectively. The amount of plural scattering
present in the low energy loss region has been used to calculate
the film thickness and is quoted above as a fraction of the
inelastic mean free path $\lambda$.
\newpage
%
\newpage
\newpage
\begin{figure}[tbp]
\begin{center}
\rotatebox{270}{\resizebox{!}{12.5cm}{\includegraphics{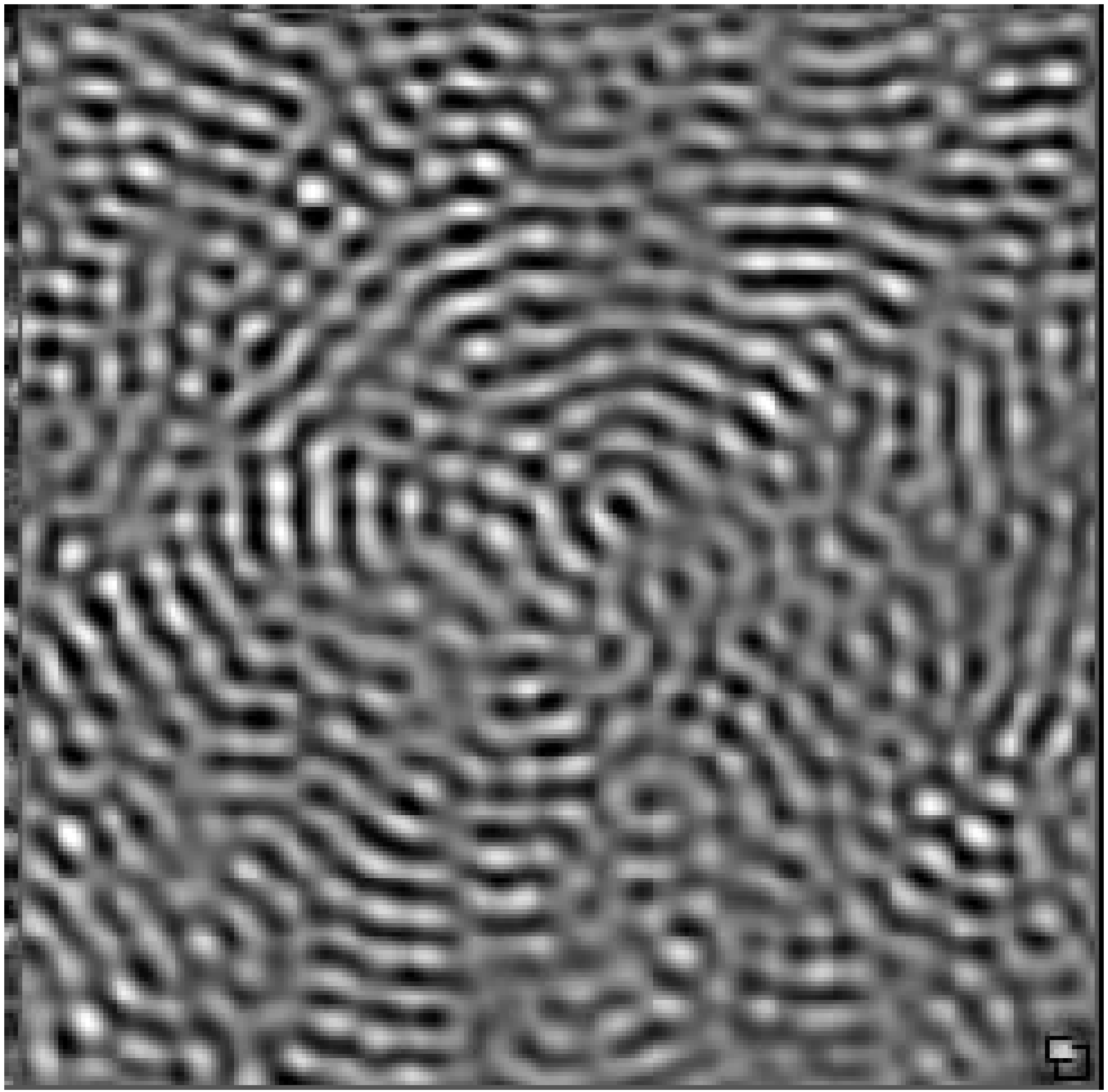}}}
\end{center}
\vspace{1cm}
\begin{center}
{\bf Figure 2(c)}
\end{center}
\vspace{10cm}
\end{figure}
\newpage
\end{document}